\documentclass[final,3p,times]{elsarticle}

\usepackage{amssymb, amsmath}
\usepackage{amsthm}
\usepackage{booktabs}
\usepackage{tabularx}
\usepackage{array}
\usepackage{siunitx}
\usepackage{algorithmic}
\usepackage{algorithm}
\usepackage{graphicx}
\usepackage{subcaption}
\usepackage[pagewise]{lineno}
\usepackage[normalem]{ulem}
\usepackage{tikz}
\usetikzlibrary{quantikz2}
\usepackage{xcolor}
\usepackage{listings}

\lstdefinestyle{pystyle}{
    language=Python,
    basicstyle=\footnotesize\ttfamily,
    keywordstyle=\color{blue},
    stringstyle=\color{green!60!black},
    commentstyle=\itshape\color{purple!40!black},
    showstringspaces=false,
    breaklines=true,
    breakatwhitespace=true,
    tabsize=4,
    frame=single,
    frameround=fttt,
    backgroundcolor=\color{gray!10},
    captionpos=b,
    morekeywords={self}
}
\lstnewenvironment{py}[1][]{\lstset{style=pystyle,#1}}{}

\tikzset{
vir/.style={circle, draw, fill=white, minimum size=4pt, inner sep=0pt}
}
\usepackage[unicode]{hyperref}
\hypersetup{
    colorlinks=true,
    linkcolor=blue,
    urlcolor=blue
}
\newcommand{\B}[1]{\boldsymbol{#1}}
\newcommand{\C}[1]{\mathcal{#1}}

\usepackage[utf8]{inputenc}
\usepackage{nomencl}
\renewcommand{\nomgroup}[1]{%
	\item[\textbf{%
		\ifthenelse{\equal{#1}{A}}{Abbreviations}{}%
		\ifthenelse{\equal{#1}{S}}{Symbols}{}
	}]%
}
\makenomenclature

\newcounter{nomseq}
\newcommand{\nomseqkey}[1]{\ifnum#1<100 0\fi\ifnum#1<10 0\fi\number#1}
\newcommand{\NOMCLA}[2]{%
	\stepcounter{nomseq}%
	\edef\nomkey{A\nomseqkey{\value{nomseq}}}%
	\expandafter\nomenclature\expandafter[\nomkey]{\(#1\)}{#2}}
\newcommand{\NOMCLF}[2]{%
	\stepcounter{nomseq}%
	\edef\nomkey{S\nomseqkey{\value{nomseq}}}%
	\expandafter\nomenclature\expandafter[\nomkey]{#1}{#2}}

\journal{XXX}

\begin{document}

\begin{frontmatter}

\title{PyQLAS: A Quantum-Classical Hybrid Solver Library for Linear Systems}

\cortext[cor]{Corresponding author}
\author[label1]{Meng-Han~Dou}
\author[label1]{Teng-Yang~Ma}
\author[label1]{Chuang-Chao~Ye}
\author[label2]{Zhao-Yun~Chen\corref{cor}}
\ead{chenzhaoyun@iai.ustc.edu.cn}
\author[label1]{Guo-Ping~Guo\corref{cor}}
\ead{dmh@originqc.com}

\affiliation[label1]{organization={Origin Quantum Computing Technology (Hefei) Co., Ltd.},
             city={Hefei},
             country={China}}
\affiliation[label2]{organization={nstitute of Artificial Intelligence, Hefei Comprehensive National Science Center},
             city={Hefei},
             country={China}}

\begin{abstract}
Quantum algorithms solve a linear system at a cost that grows only polylogarithmically in its dimension, but on present-day hardware that advantage is realised only if the quantum and the classical parts of a solve are apportioned deliberately. We present PyQLAS, a quantum--classical hybrid library for linear systems that exposes this apportionment as a design parameter rather than a fixed assumption. PyQLAS builds three nested solvers around a single variational core---a variational solver, an outer residual iteration, and a Krylov-subspace projection---so that the quantum subroutine can be held at a fixed circuit width while the classical host absorbs the growth of the problem. Around them, the library supplies preconditioning at two independent levels, dimension alignment, checkpointing, and device-budget control, and a solve written against the interface runs unchanged on an exact state-vector simulator, on a local sampler, and on a real superconducting processor. The library is validated on the IEEE 118-bus power flow, a standard test system in which the linear solver is embedded in a classical Newton iteration. There, the iteration converges to the specified tolerance and reproduces the classical reference solution to within $2.3\times10^{-8}$ p.u.\ in voltage magnitude and $3.4\times10^{-6}$ degrees in voltage angle.
\end{abstract}


\end{frontmatter}

\section{Introduction}\label{sec:1}
Linear systems are the computational core of a large part of engineering practice. They arise from finite-difference, finite-volume, and finite-element discretisations of partial differential equations \cite{Smith1985,Moukalled2016,Zienkiewicz2014}, and in a simulation they are not solved once but repeatedly, inside implicit time steps, Newton iterations, and parameter sweeps \cite{Nocedal2006,Pope2000}. Their order is set by the mesh and can be large, and their conditioning is often poor because the operator being discretised is ill conditioned \cite{Golub2013}. A solver embedded in such a workflow is a component rather than a demonstration: it must accept the matrices the application produces, return a result whose accuracy can be checked against an established reference, and behave predictably when the system is difficult. Replacing the model with a reduced-order or surrogate system avoids the repetition, but it changes the problem being solved \cite{Moin1998,Kramer2024}. Repetition, scale, conditioning, and verifiability are therefore the requirements that a practical linear-solver library has to meet.

Quantum algorithms offer an asymptotic route to such systems. Since the Harrow--Hassidim--Lloyd algorithm \cite{Harrow2009}, quantum linear-system solvers have improved steadily in their dependence on precision and on the conditioning of the problem \cite{Childs2017,Costa2022,Costa2025}, but their resource requirements assume long coherent evolution and, ultimately, error correction. On present hardware, the realistic route is variational \cite{Cerezo2021,BravoPrieto2023}, whose accuracy depends on the ansatz and on its trainability \cite{Du2022,Sukin2019,Yabo2024}. Neither route answers the engineering question by itself. What a near-term application needs is a deliberate division of labour: the quantum device prepares a bounded state or estimates a bounded quantity, while a classical host drives the iteration, decides convergence, applies preconditioning, and controls the resources spent on the device. In this model the classical host is not a scaffold around the device but the environment in which the application runs, so a quantum subroutine can be embedded in an existing simulation or optimization loop rather than called as a separate experiment. This hybrid organization is a systems property rather than an algorithmic one, and it is the property on which the usability of quantum linear solvers on small, noisy processors depends \cite{Preskill2018,Jiang2025}.

Existing software leaves this property largely unaddressed. General frameworks such as Qiskit \cite{Qiskit2019} and quantum-centric supercomputing platforms \cite{QuantumCentric2024} provide circuits, execution, and job management, but they are substrates: they contain no linear solver, and the user must assemble the solver, its iteration, and its resource controls. HHL implementations are experimental, restricted to small systems, or no longer maintained \cite{QiskitHHL,BraketHHL,MindQuantumHHL}. Variational linear solvers are distributed mainly as single-layer prototypes, framework modules, and tutorials \cite{VQLSPrototype2024,PaddleQuantum2023,ClassiqVQLS2026,PennyLaneVQLS}; they demonstrate the method, but preconditioning, dimension alignment, checkpointing, device budgets, and backend portability remain the user's responsibility. Algorithm packages and research frameworks target a particular algorithm or application \cite{PyQPandaAlgorithm2026,QCFD2026,DVQLS2026}. The available tools are thus algorithm-centric, whereas an engineering application needs a stable component with predictable behavior and a path to real hardware.

We present PyQLAS, a productized hybrid quantum--classical library for linear systems. Its interface reflects a deliberate division of labour: the device runs a bounded variational subroutine, while the classical host owns the iteration and the resource budget. Nested solver layers add classical outer loops around one variational core; moving outward keeps the circuit at a fixed width and shifts work to the host, so the classical and quantum loads can be rebalanced elastically for the problem and device at hand. The library hides this internal machinery behind a small, stable interface and the auxiliary utilities a solve requires. Application code therefore stays unchanged whether the solve runs on a simulator or on a real processor. Validation on synthetic systems and the IEEE 118-bus power flow confirms convergence to the classical reference at fixed circuit width. The contribution is therefore a reusable engineering component, not another algorithm demonstration.

The remainder of the paper is organized as follows. Section~\ref{sec:2} describes the architecture and the interface, Section~\ref{sec:3} reports the performance characterization, Section~\ref{sec:4} presents the engineering case, and Section~\ref{sec:5} concludes with usage recommendations.

\section{Architecture and Design}\label{sec:2}

\subsection{Framework Overview}\label{sec:2-1}

\begin{figure}[htbp]
    \centering
    \includegraphics[width=\linewidth]{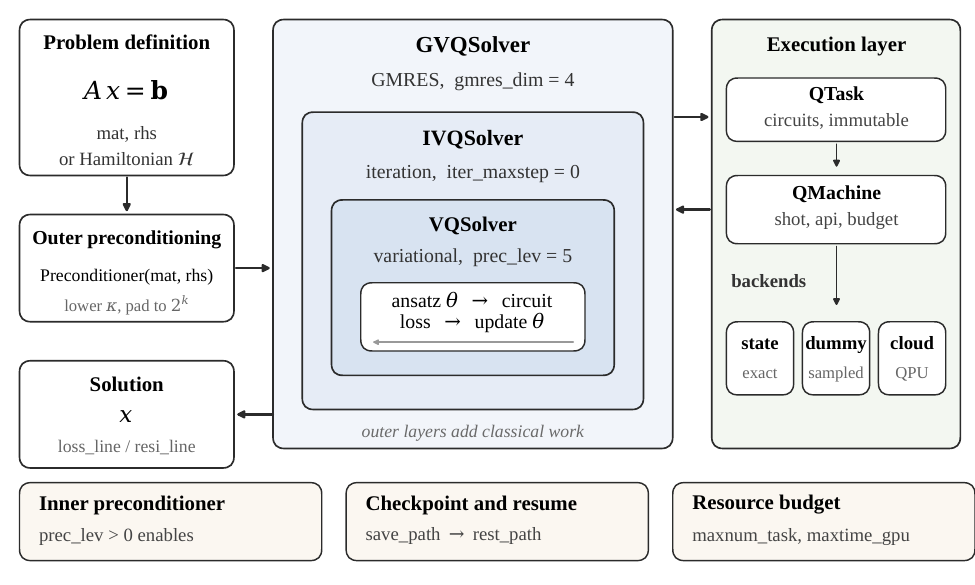}
    \caption{Overview of the PyQLAS framework. A problem is supplied as a pair of arrays or as a Hermitian operator and is passed, optionally, through the outer preconditioner, which lowers its condition number and raises its dimension to the power of two a solver requires; the transformed system is then given to one of three nested solvers. From the inside out, \texttt{VQSolver} optimizes an ansatz, \texttt{IVQSolver} adds an outer iteration on the residual, and \texttt{GVQSolver} adds a projection onto a subspace of fixed dimension; at every level the variational core repeats the same cycle of building a circuit from the current parameters, executing it, evaluating the loss, and updating the parameters. Circuits are wrapped in a \texttt{QTask} and executed by a \texttt{QMachine}, which is either an exact simulator, a local sampler, or a real processor. The three boxes at the bottom are runtime features that apply to any of the three solvers. The inner preconditioner is one of them, and is distinct from the outer preconditioner on the input path.}
    \label{fig:2-1-1}
\end{figure}

PyQLAS is a Python library for solving linear systems of equations on quantum and hybrid quantum--classical backends. It is built on \texttt{pyqpanda3} and exposes nine public entry points: an immutable task container and a stateful executor (\texttt{QTask} and \texttt{QMachine}); a variational driver (\texttt{QOptimizer}) together with three linear solvers (\texttt{VQSolver}, \texttt{IVQSolver} and \texttt{GVQSolver}); two ansatz templates (\texttt{Ansatz\_HE} and \texttt{Ansatz\_AC}); and a preprocessing function, \texttt{Preconditioner}, which acts on the system before any solver sees it. Everything else is reached through read-only properties of these objects, so the surface a user has to hold in mind stays small. The library is organized into two layers, shown in Figure~\ref{fig:2-1-1}.

Preconditioning enters at two levels, and the two are independent of one another. The \textit{outer} preconditioner is the \texttt{Preconditioner} function: it takes the system as supplied and returns a transformed one, reducing its condition number and, where necessary, raising its dimension to the power of two that a solver requires. It is entirely classical, it is applied before a solver is constructed, and it has no access to the quantum side of the computation. The \textit{inner} preconditioner belongs to the solver and is selected by the \texttt{prec\_lev} argument; it acts on the operator that the variational core minimizes, inside the solver. A solve may use either, both, or neither; the two are described together in Section~\ref{sec:2-5} and quantified separately in Sections~\ref{sec:3-3} and~\ref{sec:4}.

The solver layer turns a linear system into a sequence of parameterized quantum circuits, and owns every decision about how that sequence is driven: how many parameters the ansatz carries, when to stop, whether to precondition, and how to correct the residual. It is a ladder of three nested solvers. The base is \texttt{VQSolver}, which treats $A\B{x}=\B{b}$ as a single variational problem. \texttt{IVQSolver} wraps it in an outer iteration that corrects the residual, and \texttt{GVQSolver} wraps that layer again, projecting each outer step onto a subspace of fixed dimension. The three are independent classes rather than subclasses of one another; each simply holds a handle to the layer beneath it. The nesting is the point: as one moves outwards, the quantum device is still asked to prepare a state of the same fixed size, while the growing part of the work---iteration control, residual bookkeeping, subspace projection---is absorbed by the classical host. The variational core itself, \texttt{QOptimizer}, is exposed on its own as well, so that the same machinery can minimize an arbitrary Hermitian operator rather than only solve a system.

The execution layer is a two-object interface to a backend. \texttt{QTask} is an immutable description of quantum work: a name, one or more circuits, and, for a real device, the physical qubit block they are to occupy. \texttt{QMachine} is a stateful executor, constructed for one of three backends, which runs tasks and accumulates counters and timings. Nothing above this layer knows which backend is in use, so the same solve runs unmodified on an exact state-vector simulator, on a local sampler, or on a real processor.

Two choices cut across both layers. A solver may be handed a machine to run on, in which case several solvers can share one device and one budget; and it may be handed a directory to save into, in which case it checkpoints its state and a later solve can resume from it.

\subsection{Solution Workflow}\label{sec:2-2}

A solve begins with the problem and ends with the result, and those two ends are the only parts of it that are specific to a particular application; everything between them is fixed by the library. The problem is supplied as two NumPy arrays, \texttt{mat} and \texttt{rhs}, and it is validated at construction rather than at run time: the matrix must be square, its dimension must be a power of two, and it must not be identically zero, while the right-hand side must be one dimensional and of matching length. A rejected argument raises immediately, and the message names the offending value.

The system may then be passed, optionally, through the outer preconditioner of Section~\ref{sec:2-5}. \texttt{Preconditioner} returns a system that a solver will accept and that is cheaper to solve: an incomplete LU factorisation lowers its condition number, and the dimension is raised to the next power of two by padding with a decoupled identity block. The transformed system has the same solution as the original, and the padded entries solve to zero. A system whose dimension is already a power of two may be passed through unchanged, or through this step for the sake of its conditioning alone.

Next the outermost solver is chosen. The three solvers differ only in how much iteration they perform around the same variational core, so the choice is a question of accuracy against quantum resource consumption rather than of which algorithm to implement. Construction is where the solver derives, from \texttt{mat} and \texttt{rhs}, the operator that the variational core is to minimize, and where it fixes the number of ansatz parameters; the latter is obtained by evaluating the ansatz once, so an ansatz whose parameter count is not determined by its inputs is rejected at construction rather than in the middle of a solve. The inner preconditioning level is also a constructor argument, and it is described with the outer one in Section~\ref{sec:2-5}: it acts on the operator the variational core minimizes rather than on the supplied system, and the two settings are independent of one another.

The run itself is a repetition. Each iteration builds a circuit from the current parameters, wraps it in a \texttt{QTask}, hands it to the machine, and reads back either the state amplitudes or the sampled probabilities, depending on the backend; the loss, or the residual, is then evaluated and the parameters updated. The history of that quantity is recorded per iteration---\texttt{loss\_line} for the variational layers and \texttt{resi\_line} for the iterative ones---together with a step counter whose name likewise differs by layer (\texttt{train\_step}, \texttt{iter\_step} or \texttt{gmres\_step}). Each layer has both a tolerance and an optional step limit, and a solve stops at whichever is reached first; both defaults are permissive, the step limit being off, so that by default only the tolerance decides when a layer stops. The two are configured separately for each layer, so the inner and outer stopping rules are set independently of one another. The solution is finally available as \texttt{x}, the recorded histories are available for post-hoc analysis, and the call itself returns a status code that distinguishes a solve that converged from one that stopped because progress stalled or because a step limit was reached. When the outer preconditioner has padded the system, only the leading $n$ entries belong to the original problem; the remainder are identically zero.

If a save directory was supplied, the solver's state is checkpointed to disk as accepted steps or rounds complete, and a subsequent construction may be pointed at that file in order to continue where the previous solve stopped.

Code~\ref{code:2-1} shows the workflow end to end for a small system, solved on the exact simulator. Only the two lines that define \texttt{mat} and \texttt{rhs} are specific to the problem; exchanging \texttt{GVQSolver} for \texttt{IVQSolver} or \texttt{VQSolver}, or the ``state'' backend for one of the other two, changes nothing else in the listing.

\begin{py}[caption={A complete solve on the exact simulator.},label={code:2-1}]
import os
import numpy as np
from PyQLAS import GVQSolver, QMachine

os.makedirs('output', exist_ok=True)   # the save directory must already exist

# 8 x 8 tridiagonal system
mat = np.diag(2.0 * np.ones(8)) + np.diag(np.ones(7), 1) + np.diag(np.ones(7), -1)
rhs = np.ones(8) / np.sqrt(8)

solver = GVQSolver(mat=mat, rhs=rhs, name='solve', gmres_dim=4, prec_lev=5,
                   machine=QMachine(name='sim', type='state'),
                   save_path='output')
status = solver.Run()

residual = np.linalg.norm(rhs - mat @ solver.x) / np.linalg.norm(rhs)
print(status, solver.gmres_step, residual)
# -> 0 2 2.3606250193470965e-07
\end{py}

\subsection{Solver Layers and Default Parameters}\label{sec:2-3}

Table~\ref{tab:2-2} summarizes the four entry points. The three linear solvers share their entire tail of constructor parameters---the ansatz, the inner preconditioning level, the training limits, and the machine and checkpoint arguments---and differ in which additional iteration they introduce and in the names they give to their convergence record. A child layer is readable from its parent and is \texttt{None} once the parent has converged and has no further subproblem to solve. The outer preconditioner is not among them: it is a function rather than a solver, it holds no state, and it is applied to the system before any solver exists.

\begin{table}[htbp]
\centering
\caption{The four solver entry points. The three linear solvers accept the same tail of constructor parameters; the rows differ in the iteration they add and in the names of their convergence record and progress counter.}
\label{tab:2-2}
\begin{tabularx}{\linewidth}{@{} l l l l X @{}}
\toprule
Entry point & Problem & Sub-solver & Record / counter & Additional stopping criteria \\
\midrule
\texttt{QOptimizer} & $\C{H}$      & ---         & \texttt{loss\_line} / \texttt{train\_step} & \texttt{train\_maxstep}, \texttt{train\_conv} \\
\texttt{VQSolver}   & \texttt{mat}, \texttt{rhs} & --- & \texttt{loss\_line} / \texttt{train\_step} & \texttt{train\_maxstep}, \texttt{train\_conv} \\
\texttt{IVQSolver}  & \texttt{mat}, \texttt{rhs} & \texttt{VQSolver}  & \texttt{resi\_line} / \texttt{iter\_step}  & \texttt{iter\_maxstep}, \texttt{iter\_conv} \\
\texttt{GVQSolver}  & \texttt{mat}, \texttt{rhs} & \texttt{IVQSolver} & \texttt{resi\_line} / \texttt{gmres\_step} & \texttt{gmres\_maxstep}, \texttt{gmres\_conv} \\
\bottomrule
\end{tabularx}
\end{table}

Table~\ref{tab:2-3} lists the defaults that a user is most likely to change. Two of them deserve comment. The ansatz defaults to \texttt{Ansatz\_HE}, a hardware-efficient template whose parameter count is twice the number of qubits; the alternative, \texttt{Ansatz\_AC}, carries a single parameter but requires the inner preconditioner to be active, and falls back to the behaviour of the default template, with a warning, if it is not. The parameter-difference argument \texttt{para\_diff} selects how the gradient is obtained: a positive value, which is the default, uses a finite difference of that step size, whereas zero requests an exact gradient, which is only available for ansatz templates that support it. Supplying zero together with \texttt{Ansatz\_AC} is rejected at construction.

\begin{table}[htbp]
\centering
\caption{Selected constructor parameters, grouped by the layer that introduces each one. The three linear solvers form a stack: each accepts every parameter of the layers below it and adds its own, so a parameter is listed once, under the layer where it first appears. \texttt{QOptimizer} accepts the \texttt{VQSolver} set apart from \texttt{prec\_lev}.}
\label{tab:2-3}
\begin{tabularx}{\linewidth}{@{} l l X @{}}
\toprule
Parameter & Default & Meaning \\
\midrule
\multicolumn{3}{@{}l}{\textbf{\texttt{VQSolver}}}\\
\texttt{ansa\_func}     & \texttt{Ansatz\_HE} & Ansatz template; determines the parameter count \\
\texttt{train\_conv}    & $1\times10^{-3}$    & Convergence tolerance of the variational training \\
\texttt{train\_maxstep} & 0                   & Step limit of the variational training; \texttt{0} means no limit \\
\texttt{para\_diff}     & 0.1                 & Finite-difference step; \texttt{0} requests an exact gradient \\
\texttt{prec\_lev}      & 5                   & Inner preconditioning level; a larger value is stronger, \texttt{0} disables it, a negative value reports failure \\
\texttt{save\_path}     & empty               & Directory for the checkpoint file; must already exist when set \\
\midrule
\multicolumn{3}{@{}l}{\textbf{\texttt{IVQSolver}} adds}\\
\texttt{iter\_conv}     & $1\times10^{-6}$    & Convergence tolerance of the outer iteration \\
\texttt{iter\_maxstep}  & 0                   & Step limit of the outer iteration; \texttt{0} means no limit \\
\midrule
\multicolumn{3}{@{}l}{\textbf{\texttt{GVQSolver}} adds}\\
\texttt{gmres\_dim}     & 4                   & Subspace dimension; a power of two \\
\texttt{gmres\_aug}     & 0                   & Stabilization parameter, $0\le\texttt{gmres\_aug}<\texttt{gmres\_dim}$ \\
\texttt{gmres\_conv}    & $1\times10^{-6}$    & Convergence tolerance of the GMRES layer \\
\texttt{gmres\_maxstep} & 0                   & Step limit of the GMRES layer; \texttt{0} means no limit \\
\bottomrule
\end{tabularx}
\end{table}

All three step limits \texttt{train\_maxstep}, \texttt{iter\_maxstep} and \texttt{gmres\_maxstep} default to zero, which is not a limit of zero steps but the absence of a limit: a layer runs until its tolerance is met or until it can no longer improve. The tolerances therefore carry the whole burden of stopping by default, and a solve in which they are set too tightly will simply keep iterating; the three limits are the explicit way to put a ceiling back.

\subsection{Execution Backends and Runtime Features}\label{sec:2-4}

A \texttt{QMachine} is constructed for exactly one of three backends. The \texttt{state} backend runs on the classical host and returns the exact state vector as complex amplitudes; the \texttt{dummy} backend also runs on the host but samples each circuit, returning probabilities; and the \texttt{cloud} backend submits the same circuits unchanged to a real superconducting processor through \href{https://qcloud.originqc.com.cn/zh}{OriginQ}'s quantum cloud service, likewise returning probabilities. The choice changes only what a task returns, not how a solve is expressed, so a solve can be developed and debugged on an exact simulator, its behaviour under sampling examined locally, and the same code then submitted unchanged to hardware.

Two of the three runtime features that apply to any solver are described here; the third, preconditioning, is a mechanism in its own right with two independent levels, and it is treated in Section~\ref{sec:2-5}. Checkpointing writes a configuration file beneath \texttt{save\_path} at construction and after a run; passing that file back as \texttt{rest\_path} resumes the solve, with the stored configuration taking precedence over the arguments of the resuming call. Budget control places limits on a machine, which is what makes unattended runs on a real device safe: a task-count limit refuses further submissions before they consume anything, whereas a device-time limit can only be reported after the fact, since the cost of a task is not known in advance, and in that case the result has already been stored and is not lost.

Finally, two properties of the interface deserve to be stated explicitly, because they shape the code written around the library. Solver and machine members are read-only, so a solve in progress cannot be perturbed from outside; the single exception is a machine's \texttt{Set} method, which is the only supported way to change a machine parameter and which re-validates and rolls back on failure. And the ansatz is an extension point rather than a fixed component: a user-supplied function that follows the documented calling convention can be passed in place of either built-in template, which allows a problem-specific circuit to be used without leaving the framework.

\subsection{Preconditioning}\label{sec:2-5}

Section~\ref{sec:2-1} introduced the two levels of preconditioning; this section describes them. Because the two share a name, they are worth separating explicitly before either is set out in detail. The \textit{inner} preconditioner belongs to a solver: it is selected at construction by an integer argument and it acts on the operator that the variational core minimizes. The \textit{outer} preconditioner is a function that the user calls before a solver exists: it is entirely classical, it holds no state, and it acts on the system supplied to the solver. Neither changes the other's setting, and a solve may use either, both, or neither; Table~\ref{tab:2-4} places the two side by side.

\begin{table}[htbp]
\centering
\caption{The two levels of preconditioning in PyQLAS\@. They are configured independently of one another, and a solve may use either, both, or neither.}
\label{tab:2-4}
\begin{tabularx}{\linewidth}{@{} l X X @{}}
\toprule
 & Inner preconditioner & Outer preconditioner \\
\midrule
Selected by  & \texttt{prec\_lev}, an integer constructor argument of the solver & the function \texttt{Preconditioner}, called on the system by the user \\
Acts on      & the operator that the variational core minimizes & the system supplied to the solver \\
When it runs & inside the solver, at every iteration of the variational core & once, before any solver is constructed \\
What it buys & convergence: it extends the range of condition numbers over which a solve succeeds & admissibility and economy: it raises the dimension to a power of two and lowers the condition number \\
\bottomrule
\end{tabularx}
\end{table}

The \textit{inner} preconditioner is the one internal to a solver, and it is controlled by a single integer, \texttt{prec\_lev}: values above zero enable it and larger values request a stronger preconditioner, zero disables it, and a negative value reports that preconditioning was requested but could not be achieved. The level actually in force is readable after construction, so a solve can be audited rather than assumed. Because it acts on the operator the variational core sees, it addresses a failure internal to the iteration rather than a property of the problem: it cannot make an inadmissible system admissible, and it has nothing to say about a system's conditioning before the first parameter update. Its effect is quantified in Section~\ref{sec:3-3}, where it is the mechanism that extends the range of condition numbers a solve survives, and in Section~\ref{sec:4}, where it carries the convergence of the engineering case.

The \textit{outer} preconditioner is the function \texttt{Preconditioner}. It addresses the two properties of a system that are settled before any iteration begins, and that the inner one therefore cannot reach: a dimension that is not a power of two, which a solver refuses outright, and a condition number that is large from the start. It is entirely classical and is applied to the system before a solver is constructed. It returns a system that has the same solution as the one it was given, and three operations are involved, of which only the first changes the conditioning: an incomplete LU factorisation is applied on the left, both sides are then rescaled by the median singular value of the factorised matrix, and finally the dimension is raised to the next power of two by padding, the added entries of the right-hand side being set to zero. The factorisation is selected by a \texttt{method} argument: \texttt{method='ilu'}, the default, performs all three operations, whereas \texttt{method='none'} omits the factorisation and keeps the rescaling and the padding alone, so that the system is made admissible without being better conditioned. The two settings form the controlled pair used in Section~\ref{sec:4}.

\begin{py}[caption={Preparing a system with the outer preconditioner.},label={code:2-2}]
PA, PB = Preconditioner(mat, rhs)      # same solution, lower kappa, order 2^k

solver = GVQSolver(mat=PA, rhs=PB, name='solve', gmres_dim=4, prec_lev=5,
                   machine=QMachine(name='sim', type='state'),
                   save_path='output')
status = solver.Run()

x = solver.x[:len(rhs)]                # the leading n entries are the solution
\end{py}

Padding with an identity block rather than zeros matters. A zero block would make the matrix singular, and the system would no longer have a unique solution; an identity block leaves the added rows fully decoupled, so that the extra entries solve to zero and the leading $n$ entries of the solution are those of the original problem. The same choice leaves the spectrum of the original matrix untouched, which a sharper rescaling would not.

The padded dimension is the smallest power of two not below the original, so a system of order $181$ becomes one of order $256$ while a system already of order $1024$ is passed through unchanged. The factorisation is classical and, at the sizes a variational solver can reach, cheap; what the function buys is measured in Section~\ref{sec:4}. Its strength is set by the drop tolerance and fill-in limit of the factorisation, and the two directions of error are not symmetric: a tolerance set too loosely discards so much of the matrix that the result can be worse conditioned than the system that went in.

\section{Performance Testing}\label{sec:3}

\subsection{Experimental setup}\label{sec:3-1}

All experiments in this work are performed on the three backends defined in Section~\ref{sec:2-4}. The classical host on which the ``state'' and ``dummy'' backends run is described in Table~\ref{tab:classical_params}, and the ``cloud'' backend executes on the OriginQ ``WK\_C180'' superconducting quantum computer, whose parameters are given in Table~\ref{tab:quantum_params}.

\begin{table}[H]
\centering
\caption{Parameters of the classical host computer.}
\label{tab:classical_params}
\begin{tabularx}{\linewidth}{@{} l @{\hspace{16pt}} X @{\hspace{26pt}} l @{\hspace{16pt}} X @{}}
\toprule
Item & Specification & Item & Specification \\
\midrule
CPU       & Intel Core i5-10500 @ \SI{3.10}{GHz} & CPU cores & 6 \\
Memory    & \SI{32.0}{GB}                        & Storage   & \SI{512}{GB} SSD + \SI{2}{TB} HDD \\
OS        & Windows 10, 64-bit                   & Graphics  & Intel UHD Graphics 630, \SI{128}{MB} \\
\bottomrule
\end{tabularx}
\end{table}

\begin{table}[H]
\centering
\caption{Parameters of the OriginQ WK\_C180 superconducting quantum processor.}
\label{tab:quantum_params}
\begin{tabularx}{\linewidth}{@{} l @{\hspace{20pt}} X @{\hspace{26pt}} l @{\hspace{20pt}} X @{}}
\toprule
Item & Specification & Item & Specification \\
\midrule
$T_1$ (avg)      & \SI{35.31}{\micro\second} & $T_2$ (avg)  & \SI{13.78}{\micro\second} \\
1Q fidelity      & 0.9988                    & 2Q fidelity  & 0.9790 \\
Readout fidelity & 0.9260                    & Native gates & $R_\phi$, $CZ$ \\
Max depth        & 500                       & Max shots    & 49600 \\
\bottomrule
\end{tabularx}
\end{table}

\subsection{Performance vs. Number of Qubits}\label{sec:3-2}

This subsection evaluates the performance of PyQLAS as a function of the number of qubits $n$. The number of qubits is fixed by the matrix dimension through $n=\log_2 N$, and it characterizes the size of the linear system; the figures below are therefore all plotted against $n$. The coefficient matrix $A$ is fixed to the identity matrix and the right-hand side $\B{b}$ is a random normalized real vector. The ansatz is a hardware-efficient ansatz (HEA) consisting of two $R_Y$ layers and one $\mathrm{CNOT}$ layer, giving a total of $2n$ ansatz parameters. All runs use the ``state'' backend. Throughout this subsection the convergence tolerances are held fixed at \texttt{train\_conv}$=1\times10^{-4}$ for the VQSolver layer, \texttt{iter\_conv}$=1\times10^{-2}$ for the IVQSolver layer and \texttt{gmres\_conv}$=1\times10^{-2}$ for the GVQSolver layer.

\begin{figure}[htbp]
    \centering
    \includegraphics[width=\linewidth]{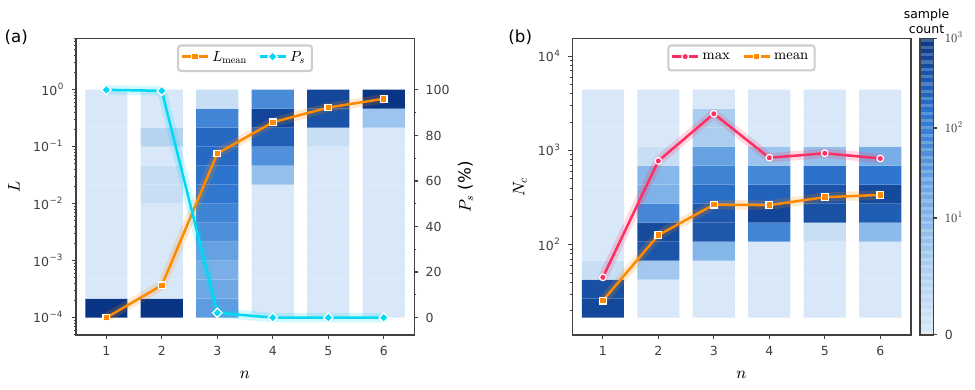}
    \caption{VQSolver performance versus the number of qubits $n$, over 1000 random right-hand sides per $n$. (a) Distribution of the final loss $L$ (stacked bars), with the mean loss $L_{\mathrm{mean}}$ (line, left axis) and the success ratio $P_s$ (line, right axis). (b) Distribution of the number of circuits $N_c$ (stacked bars), with its maximum and mean (lines). In both panels the color encodes the number of samples.}
    \label{fig:3-2-1}
\end{figure}

Figure~\ref{fig:3-2-1} shows the performance of VQSolver as a function of the number of qubits. Each value of $n$ is tested on 1000 random right-hand sides, and a training run is counted as successful when the final loss satisfies $L\le1\times10^{-4}$. Figure~\ref{fig:3-2-1}(a) presents the loss distribution and the resulting success ratio $P_s$. For $n\le 2$ the success ratio is essentially 100\%, and the loss sits well below the threshold rather than merely meeting it: the median loss is $1.3\times10^{-10}$ at $n=1$ and $9.5\times10^{-6}$ at $n=2$, the latter just under the $10^{-4}$ by which a run is judged. At $n=3$ the distribution broadens markedly, spanning about four orders of magnitude, and the success ratio falls abruptly to 2.1\%. For $n\ge4$ the solver fails on every sample: the success ratio is zero at $n=4,5,6$.

\begin{figure}[htbp]
    \centering
    \includegraphics[width=\linewidth]{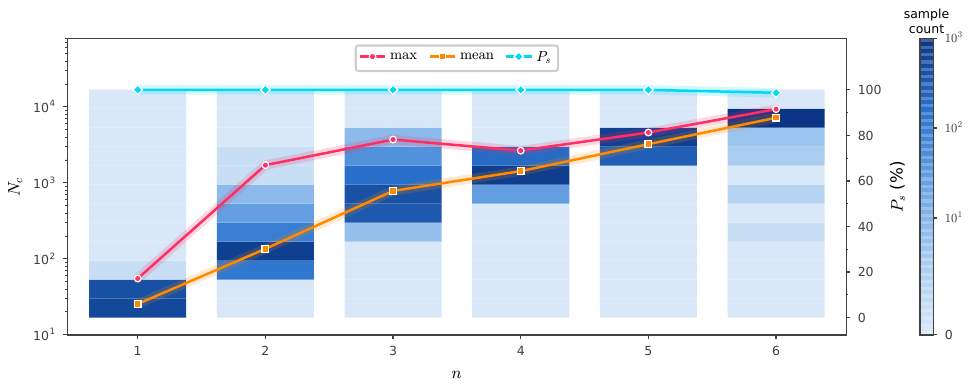}
    \caption{IVQSolver performance versus the number of qubits $n$, over 1000 random right-hand sides per $n$. Distribution of the number of circuits $N_c$ (stacked bars), with its maximum and mean (lines, left axis) and the success ratio $P_s$ (line, right axis). The color encodes the number of samples.}
    \label{fig:3-2-2}
\end{figure}

Figure~\ref{fig:3-2-1}(b) shows the number of quantum circuits $N_c$ consumed by VQSolver. For $n\le 2$, $N_c$ grows with $n$, indicating that successful training requires more circuits as the problem grows. As $n$ increases further, this correlation breaks down, because the success ratio can no longer be guaranteed. Once only a few percent of the runs reach the target loss, the number of circuits in a single run is governed by the stopping rule of the optimizer rather than by convergence, so that the values are no longer comparable across different $n$.

Figure~\ref{fig:3-2-2} shows the performance of IVQSolver for different numbers of qubits. Since $A=I$ and $\|\B{b}\|_2=1$ imply $r=\|\B{b}-A\B{x}\|_2=\sqrt{L}$, a solution is counted as successfully obtained when $r\le 1\times10^{-2}$, which is exactly the same criterion as the $L\le1\times10^{-4}$ threshold used for VQSolver above; the two figures are therefore directly comparable. IVQSolver does not require a high success ratio for its internal training, because the accuracy of the final solution is guaranteed by the iterative mechanism. Accordingly, the success ratio of IVQSolver remains almost 100\% throughout, with only a very small number of samples failing to be solved at $n=6$. The mean curve of $N_c$ shows that, regardless of the internal training success ratio, the number of circuits consumed by IVQSolver grows exponentially with the number of qubits under the same convergence condition.

\begin{figure}[htbp]
    \centering
    \includegraphics[width=\linewidth]{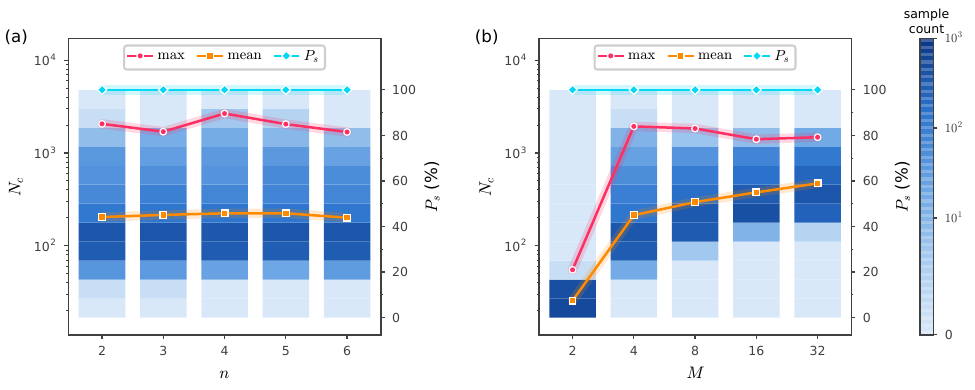}
    \caption{GVQSolver performance versus the number of qubits with the subspace dimension fixed at $M=4$ (a), and versus the subspace dimension with the matrix dimension fixed at $N=32$ (b); 1000 random right-hand sides per setting. Distribution of the number of circuits $N_c$ (stacked bars), with the maximum and the mean (lines, left axis) and the success ratio $P_s$ (line, right axis). The color encodes the number of samples.}
    \label{fig:3-2-3}
\end{figure}

Figure~\ref{fig:3-2-3} shows the performance of GVQSolver. In Figure~\ref{fig:3-2-3}(a), the subspace dimension is fixed at $M=4$, so that only $\log_2 M=2$ qubits are actually used; consequently, the number of quantum circuits consumed by GVQSolver does not change appreciably as the number of qubits increases, which implies that the computational cost of solving the linear system is shifted further toward the classical computer. Figure~\ref{fig:3-2-3}(b) presents the results obtained by fixing the matrix dimension at $N=32$ and varying the subspace dimension $M$. As $M$ increases, the average number of circuits consumed grows.

Compared with VQSolver, GVQSolver retains the convergence guaranteed by the iterative strategy, as demonstrated by the fact that all samples in Figure~\ref{fig:3-2-3} are successfully solved. In contrast to IVQSolver, which reaches the same success ratio through outer residual refinement at the cost of a number of circuits that grows with the number of qubits, GVQSolver confines the quantum work to a Krylov subspace of fixed dimension and absorbs the growth in the classical solver. By incorporating the GMRES method, GVQSolver provides a flexible means of tuning quantum and classical resources, thereby enabling the solution of larger-scale problems under limited quantum resources.

\subsection{Performance vs. Matrix Condition Number}\label{sec:3-3}
This subsection isolates the effect of the condition number $\kappa$ on the three solvers. The test matrices are drawn from the normalized tridiagonal family
\begin{equation}
    A = -\frac{1}{a+c}\sum_{i=1}^{N-1}\ket{i}\bra{i-1}
      + \frac{a}{a+c}\sum_{i=0}^{N-1}\ket{i}\bra{i}
      - \frac{1}{a+c}\sum_{i=0}^{N-2}\ket{i}\bra{i+1},
\end{equation}
where $N=4$, $c=2\cos(\pi/5)$ and $a=c(\kappa+1)/(\kappa-1)$. The condition number of the matrix $A$ constructed in this way is exactly $\kappa$. The right-hand side $\B{b}$ is again a normalized nonzero real vector, and each value of $\kappa$ is tested on 1000 random samples. Unless stated otherwise, the solver tolerances are \texttt{train\_conv}$=1\times10^{-4}$, \texttt{iter\_conv}$=1\times10^{-2}$ and \texttt{gmres\_conv}$=1\times10^{-2}$. Apart from these tolerances, $\kappa$ is the only quantity that varies across the sweep.

\begin{figure}[htbp]
    \centering
    \includegraphics[width=\linewidth]{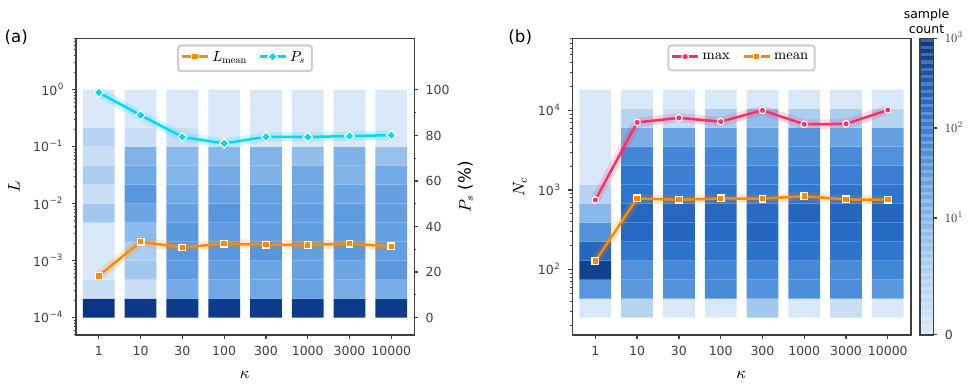}
    \caption{VQSolver performance versus matrix condition number, over 1000 random right-hand sides per $\kappa$. (a) Distribution of the final loss $L$ (stacked bars), with the mean loss $L_{\mathrm{mean}}$ (line, left axis) and the success ratio $P_s$ (line, right axis). (b) Distribution of the number of circuits $N_c$ (stacked bars), with its maximum and mean (lines). In both panels the color encodes the number of samples.}
    \label{fig:3-3-1}
\end{figure}

Figure~\ref{fig:3-3-1} shows how the performance of VQSolver varies with the condition number. A run is counted as a successful training when the final loss satisfies $L\le1\times10^{-4}$. In panel (a), the success ratio $P_s$ falls from 98.8\% at $\kappa=1$ to 79.3\% at $\kappa=30$, and then remains flat. The number of circuits consumed in panel (b) exhibits a similar trend: as $\kappa$ increases, it first grows rapidly and then remains essentially unchanged. Overall, a larger $\kappa$ degrades both accuracy and efficiency. More notably, this degradation is not gradual in $\kappa$: it is essentially complete at an early stage. Counteracting it therefore requires a preconditioner strong enough to reduce the condition number of the system to a very small value.

\begin{figure}[tbp]
    \centering
    \includegraphics[width=\linewidth]{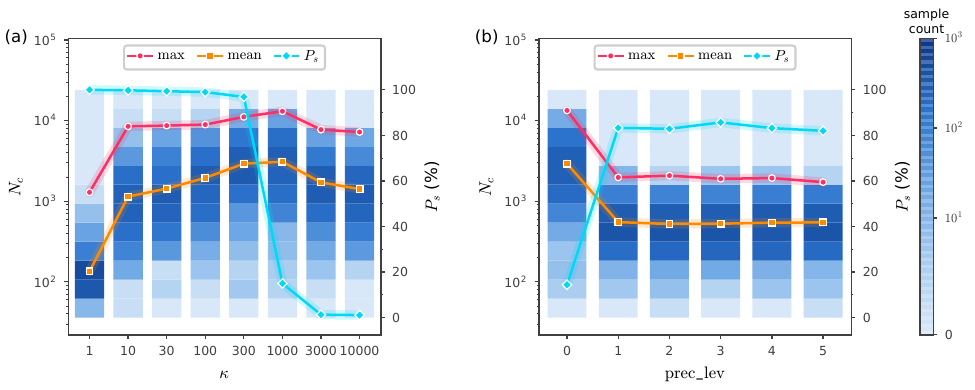}
    \caption{IVQSolver performance versus the matrix condition number (a) and the preconditioning level (b), over 1000 random right-hand sides per setting. Distribution of the number of circuits $N_c$ (stacked bars), with its maximum and mean (lines, left axis) and the success ratio $P_s$ (line, right axis). The color encodes the number of samples.}
    \label{fig:3-3-2}
\end{figure}

Figure~\ref{fig:3-3-2} shows the effect of the condition number on IVQSolver. A solution is counted as successfully obtained when the residual satisfies $r\le1\times10^{-2}$. Panel (a) sweeps $\kappa$ with preconditioning disabled. The success ratio remains almost 100\% up to $\kappa=300$; as $\kappa$ increases further, it falls sharply to nearly 0\%. The mean number of circuits falls at the two largest values of $\kappa$, but this reflects the fact that 98.7\% and 98.9\% of the runs terminate early there, once the residual can no longer be improved, rather than a genuine saving in cost. These results indicate that IVQSolver can mitigate the effect of $\kappa$ on accuracy within a certain range, but cannot eliminate it.

Figure~\ref{fig:3-3-2}(b) examines the effect of the inner preconditioner built into PyQLAS at $\kappa=1000$. The parameter \texttt{prec\_lev} controls it: larger values correspond to a stronger preconditioner, and \texttt{prec\_lev}$=0$ disables it. Enabling it (\texttt{prec\_lev}$>0$) raises the success ratio from 15\% to approximately 80\% while reducing the number of circuits consumed by about one order of magnitude.

\begin{figure}[htbp]
    \centering
    \includegraphics[width=\linewidth]{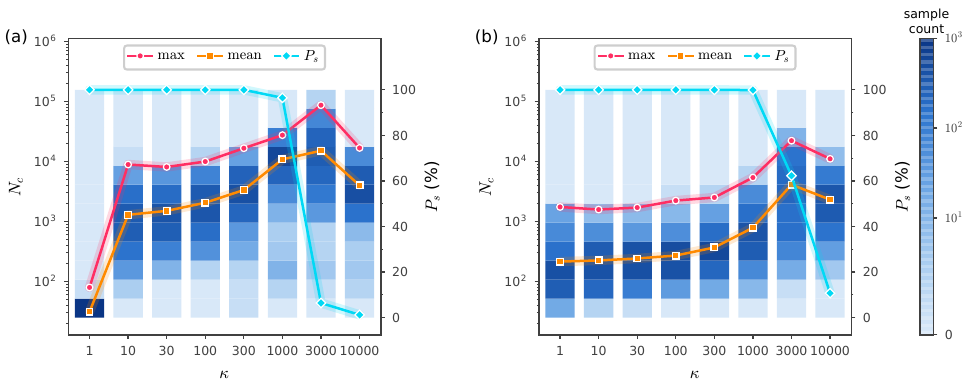}
    \caption{GVQSolver performance versus matrix condition number, over 1000 random right-hand sides per $\kappa$. (a) Inner preconditioner off. (b) Library default. Distribution of the number of circuits $N_c$ (stacked bars), with the maximum and the mean (lines, left axis) and the success ratio $P_s$ (line, right axis). The color encodes the number of samples.}
    \label{fig:3-3-3}
\end{figure}

Figure~\ref{fig:3-3-3} shows GVQSolver on the same grid, with the two panels differing only in whether the inner preconditioner is enabled. The convergence tolerance of the inner IVQSolver layer is \texttt{iter\_conv}$=1\times10^{-2}$, and the criterion for a successful solution remains $r\le1\times10^{-2}$. The subspace dimension is $M=4$. The trend in Figure~\ref{fig:3-3-3}(a) is qualitatively consistent with that in Figure~\ref{fig:3-3-2}(a). GVQSolver extends the range of condition numbers that can be tolerated, but only by a single step of the $\kappa$ grid, from $\kappa=300$ to $\kappa=1000$: at $\kappa=1000$ it still solves 96.5\% of the samples, whereas IVQSolver manages only 15.0\%, while beyond that point both collapse.

Figure~\ref{fig:3-3-3}(b) shows the performance of GVQSolver for different condition numbers when the inner preconditioner is enabled (\texttt{prec\_lev}$=5$). As in Figure~\ref{fig:3-3-2}(b), enabling it reduces $N_c$, confirming that it saves quantum resources for both IVQSolver and GVQSolver. One anomaly remains, however: beyond $\kappa=1000$ the success ratio of GVQSolver falls rapidly, indicating that the convergence guarantee provided by the iterative strategy breaks down.

\begin{figure}[tbp]
    \centering
    \includegraphics[width=\linewidth]{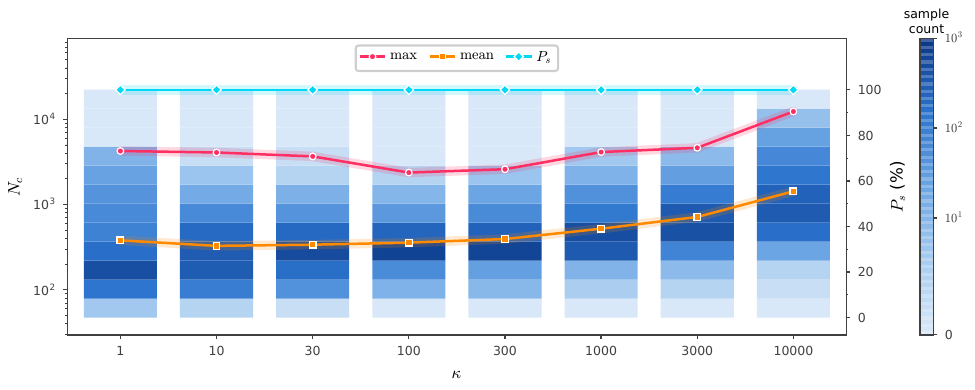}
    \caption{GVQSolver performance versus matrix condition number with tightened inner tolerances, over 1000 random right-hand sides per $\kappa$. Distribution of the number of circuits $N_c$ (stacked bars), with the maximum and the mean (lines, left axis) and the success ratio $P_s$ (line, right axis). The color encodes the number of samples.}
    \label{fig:3-3-4}
\end{figure}

To investigate this breakdown further, the tolerances of the two inner solver layers are tightened: \texttt{iter\_conv} from $1\times10^{-2}$ to $1\times10^{-3}$ and \texttt{train\_conv} from $1\times10^{-4}$ to $1\times10^{-6}$; the outer tolerance \texttt{gmres\_conv} is left unchanged at $1\times10^{-2}$. Figure~\ref{fig:3-3-4} shows the performance of GVQSolver over the same range of $\kappa$ under this setting, with \texttt{prec\_lev}$=5$ throughout. Compared with Figure~\ref{fig:3-3-3}(b), tightening the inner tolerances restores 100\% convergence at every value of $\kappa$, and does so without a significant increase in the number of circuits consumed relative to the low-accuracy setting. This indicates that tightening the convergence tolerances of the inner solver layers suffices to guarantee the success ratio of GVQSolver. A plausible explanation is that although a more accurate inner solver consumes more circuits per inner iteration, it also requires fewer iterations overall, so that the total number of circuits does not grow.

\subsection{Performance vs. Number of Ansatz Parameters}\label{sec:3-4}
This subsection examines how the size of the ansatz limits the solvers, and how far the classical layers around the variational solver compensate for that limit. The coefficient matrix is again fixed to $A=I$, now with $N=8$, so that the exact solution is $\B{b}$ itself and the number of variational parameters is the only quantity that varies. The ansatz is a family of $k$ $R_Y$ layers separated by $k-1$ CNOT ladders, containing $N_\theta=3k$ parameters; sweeping $k=1,\dots,5$ gives $N_\theta\in\{3,6,9,12,15\}$; each value is tested on 1000 random right-hand sides with the tolerances of Section~\ref{sec:3-2}.

\begin{figure}[htbp]
    \centering
    \includegraphics[width=\linewidth]{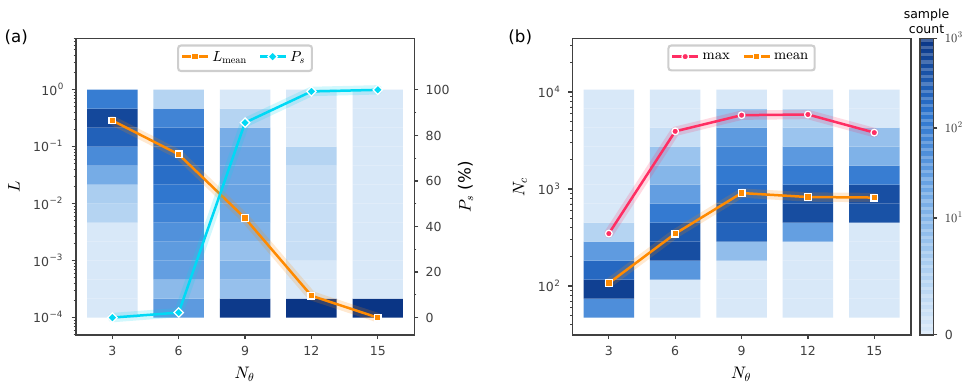}
    \caption{VQSolver performance versus the number of ansatz parameters $N_\theta$, over 1000 random right-hand sides per $N_\theta$. (a) Distribution of the final loss $L$ (stacked bars), with the mean loss $L_{\mathrm{mean}}$ (line, left axis) and the success ratio $P_s$ (line, right axis). (b) Distribution of the number of circuits $N_c$ (stacked bars), with its maximum and mean (lines). In both panels the color encodes the number of samples.}
    \label{fig:3-4-1}
\end{figure}

Because $A=I$ and $\|\B{b}\|_2=1$ make the loss and the squared residual coincide, the success criteria of the two figures below---$L\le1\times10^{-4}$ and $r\le1\times10^{-2}$---are one and the same condition, so the figures can be read against each other. The range of the scan is set by a degree-of-freedom count: a generic real state in a $2^{n}$-dimensional space requires $2^{n}-1$ real parameters, so such an ansatz cannot represent an arbitrary $\B{b}$ with fewer than 7 parameters. This is a necessary condition rather than a sufficient one, and it holds for this family only. Sweeping 3, 6, 9, 12 and 15 parameters brackets that count, placing the expected transition between $N_\theta=6$ and $N_\theta=9$.

Figure~\ref{fig:3-4-1} shows the result for VQSolver. Panel (a) displays two regimes: below the parameter threshold very few runs succeed, while above it the loss distribution collapses onto the success threshold. The success ratio jumps from 2\% at $N_\theta=6$ to 86\% at $N_\theta=9$, exactly where the number of parameters crosses $2^{n}-1=7$. Above the threshold both curves saturate: further parameters change neither the success ratio nor the cost, and the small values of $N_c$ below the threshold belong to runs that fail rather than to a cheaper solve. The sensitivity to $N_\theta$ is therefore confined to a narrow window around seven. Since $2^{n}-1$ grows exponentially whereas the number of parameters of a practical HEA grows only linearly, this suggests that at large $n$ a low-parameter HEA should not be expected to solve the system at all.

\begin{figure}[htbp]
    \centering
    \includegraphics[width=\linewidth]{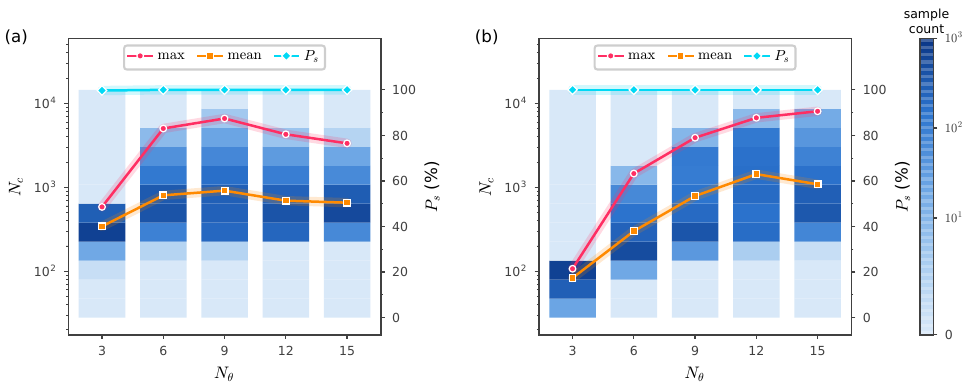}
    \caption{Outer solver layers, IVQSolver (a) and GVQSolver (b), versus the number of ansatz parameters $N_\theta$, over 1000 random right-hand sides per $N_\theta$. Distribution of the number of circuits $N_c$, with the maximum and the mean, and the success ratio $P_s$ on the right axis; the color encodes the number of samples.}
    \label{fig:3-4-2}
\end{figure}

Figure~\ref{fig:3-4-2} shows the same scan applied to the two iterative solvers, whose panels share the system, the ansatz family and the values of $N_\theta$. Both reach the target accuracy on essentially every sample at every value of $N_\theta$, including $N_\theta=3$, where VQSolver never succeeds; in both panels the mean number of circuits is also lowest at $N_\theta=3$, and for GVQSolver that minimum is about a quarter of that of IVQSolver on the same system. Particularly for GVQSolver, a small ansatz is therefore not merely adequate but also the cheapest choice.

\subsection{Performance vs. Quantum Backend}\label{sec:3-5}
This subsection concerns the cost of execution rather than the accuracy of the solution, and uses GVQSolver only. Throughout, cost is reported as \textit{machine time}: the time the \texttt{QMachine} spends running the circuits, as its own counter records it, rather than the time taken by the classical code around it. A solver may be built on any of the three backends, so what that time contains depends on which machine is in use---for the ``state'' and ``dummy'' backends it is the simulator running on the classical host, whereas for the ``cloud'' backend it is the time of a job on the real device, covering submission, queuing and retrieval as well as the interval for which the chip itself is occupied.

\begin{figure}[htbp]
    \centering
    \includegraphics[width=\linewidth]{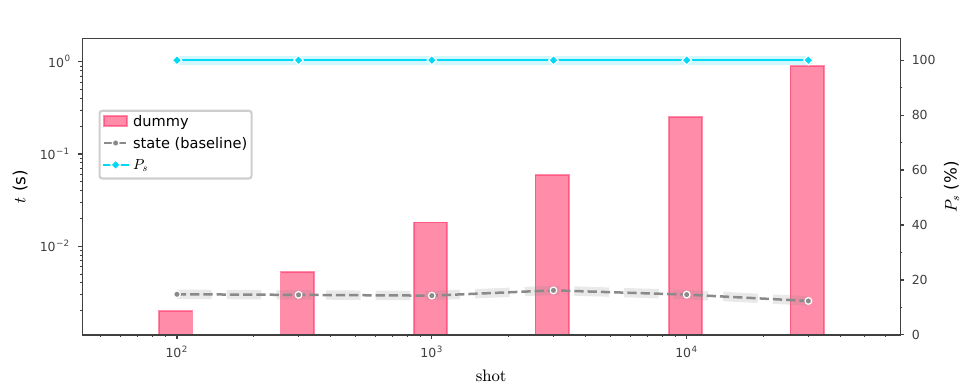}
    \caption{Machine time of a solve versus the number of shots, over 100 random right-hand sides per setting, on the ``state'' and ``dummy'' backends. Left axis, the mean time the machine spends on a solve; right axis, success ratio $P_s$.}
    \label{fig:3-5-1}
\end{figure}

\begin{figure}[htbp]
    \centering
    \includegraphics[width=\linewidth]{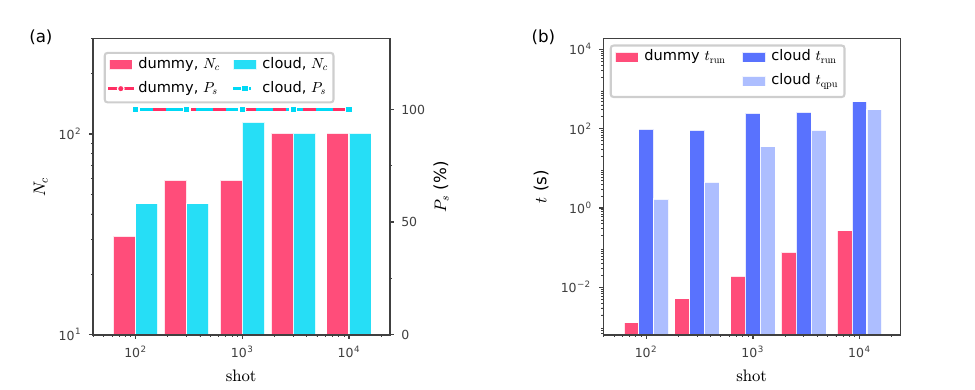}
    \caption{Comparison of the ``dummy'' and ``cloud'' backends versus the number of shots. (a) Number of circuits $N_c$, as bars, with the success ratio $P_s$ as lines on the right axis. (b) Machine time per solve, decomposed into the total run time and the QPU time for both backends. The color distinguishes the backend.}
    \label{fig:3-5-2}
\end{figure}

The matrix is $A=I_8$, with the solver settings of Section~\ref{sec:3-2} except that the subspace dimension is $M=4$. Two comparisons are made: between the two virtual backends, ``state'' and ``dummy'', in Figure~\ref{fig:3-5-1}, over 100 right-hand sides per value of the number of shots; and between the ``dummy'' and ``cloud'' backends in Figure~\ref{fig:3-5-2}, where the number of shots is swept only up to $10^{4}$, well below the $49600$ shots the ``cloud'' backend permits, and, to bound the device time, a single fixed right-hand side is used.

Figure~\ref{fig:3-5-1} shows the machine time of a single solve against the number of shots on the two virtual backends. The ``dummy'' backend samples every circuit, so its cost grows with the number of shots, whereas the ``state'' backend, which serves as the exact baseline, returns an exact state vector and is unaffected by it; the two curves therefore diverge. The growth is nearly proportional to the number of shots up to $3000$ shots and turns mildly superlinear beyond that. For the smallest numbers of shots the order is reversed, with the ``dummy'' backend the faster of the two, and the curves cross between $100$ and $300$ shots: the ``state'' backend must evaluate the whole state vector for every circuit, whereas the ``dummy'' backend still has few shots to draw. The success ratio is 100\% at every point on both backends, so this figure isolates cost and carries no information about the effect of sampling noise on the solution.

Figure~\ref{fig:3-5-2} replaces the ``state'' backend with the ``cloud'' backend. For this sample the residual is at machine precision on both backends, far below the $10^{-2}$ criterion, so the sampling noise of the ``cloud'' backend does not degrade the converged solution. The values of $N_c$ in panel (a) are likewise of the same order on the two backends. Panel (b) shows where the time goes. A hardware solve costs three to five orders of magnitude more machine time than the same solve on the ``dummy'' backend, but almost none of that is QPU time when the number of shots is small: the time the chip is occupied is under 2\% of that machine time at 100 shots and rises to about 40\% of it at $10^{4}$. Submission, queuing and retrieval around each job therefore dominate when the number of shots is small. Each point in this figure rests on a single right-hand side, so the curves provide a single realization rather than a statistical comparison.

The last result deserves a comment, because it runs counter to the naive expectation. A larger number of shots reduces the sampling error of each circuit, which one would expect to improve the accuracy of each quantum subroutine, shorten the optimization, and thereby reduce the total number of circuits. This is not what is observed: over the range studied the number of circuits consumed does not fall as the number of shots grows. A plausible reading is that, in a hybrid quantum-classical architecture such as that of the GVQSolver, sampling error does not act as a mere perturbation of an otherwise fixed trajectory. The quantity it perturbs is read back three times over: as the loss estimated by the variational core, as the residual on which the outer layer decides whether to continue, and as the basis onto which the Krylov layer projects. A noisier estimate therefore does not simply lengthen or shorten a path laid out in advance; it selects a different path, and whether that path costs more circuits or fewer becomes a property of the particular solve rather than a monotone function of the number of shots. Settling the question would require averaging over many right-hand sides at each number of shots, which the device-time budget of this study did not allow.


\section{Practical Applications in Engineering Scenarios}\label{sec:4}
This section uses the IEEE 118-bus test system. The case is useful for three
reasons: it is nonlinear, so the linear solver sits inside an outer iteration rather than
standing alone; its Jacobians are ill conditioned, the regime Section~\ref{sec:3-3}
identified as damaging to the solvers; and its solution has a physical meaning that can be
checked against a classical reference. The power flow problem asks for the state at every
bus---a voltage magnitude $V$ and a phase angle $\theta$---such that the power injected at
each bus matches the generation and demand specified there; it is a set of nonlinear
algebraic equations in $V$ and $\theta$, solved here by a classical Newton--Raphson
iteration. At each step the mismatch $\B{m}$ between the specified and the computed power
is formed, the Jacobian $\B{J}$ of that mismatch with respect to the state is assembled,
and the linear system $\B{J}\,\Delta\B{x} = \B{m}$ is solved for the correction
$\Delta\B{x}$, which is then applied to the state. Table~\ref{tab:4-1} lists the case and
the configuration used throughout.

\begin{table}[H]
\centering
\caption{The IEEE 118-bus power flow case and the configuration used to solve its Newton systems.}
\label{tab:4-1}
\begin{tabularx}{\linewidth}{@{} l @{\hspace{20pt}} X @{\hspace{26pt}} l @{\hspace{20pt}} X @{}}
\toprule
Item & Value & Item & Value \\
\midrule
Buses            & 118                          & Solver               & \texttt{GVQSolver} \\
Branches         & 186                          & Subspace dimension   & $M=4$ ($2$ qubits) \\
Slack bus        & 69                           & Inner preconditioner & \texttt{prec\_lev} $=5$ \\
PV / PQ buses    & 53 / 64                      & \texttt{iter\_conv} / \texttt{train\_conv} & $10^{-4}$ / $10^{-6}$ \\
Jacobian order   & 181                          & \texttt{gmres\_conv} & $10^{-3}$ \\
After padding    & 256                          & Backend              & \texttt{state} (exact) \\
Base power       & \SI{100}{MVA}                & Newton tolerance     & $10^{-6}$ p.u. \\
\bottomrule
\end{tabularx}
\end{table}

The Jacobian of this case is of order $181$, which is not a power of two, and it is ill conditioned, with $\kappa(\B{J})\approx 3.1\times10^{3}$ at the flat start and $\approx 3.3\times10^{3}$ once the voltages have moved. Both are handled before the solver is reached, by the outer preconditioner of Section~\ref{sec:2-5}: applied to this Jacobian it returns a system of order $256$ whose solution is the correction $\Delta\B{x}$, and for the first Newton step it takes $\kappa$ from $3.1\times10^{3}$ to $1.3\times10^{2}$, holding it between $1.3\times10^{2}$ and $1.4\times10^{2}$ across the six steps of the run.

\begin{figure}[htb]
    \centering
    \includegraphics[width=\linewidth]{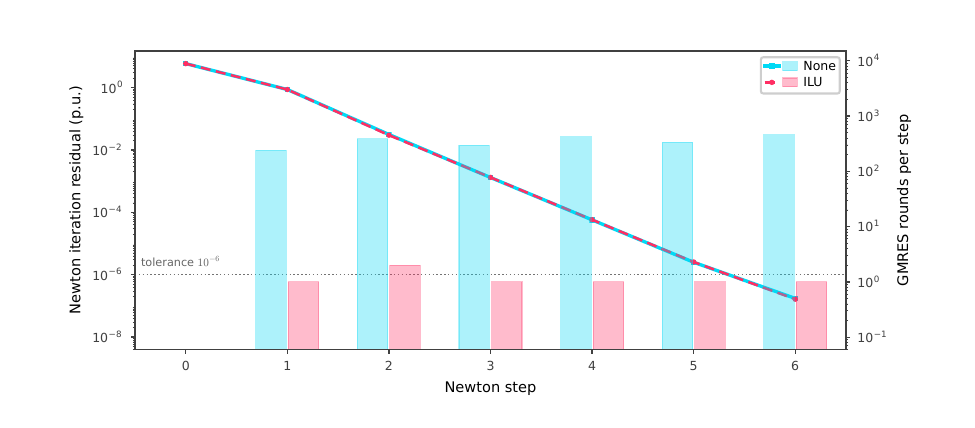}
    \caption{Newton--Raphson convergence on the IEEE 118-bus system, with and without the outer preconditioner. Left axis, the largest power mismatch at each Newton step, in per unit; right axis, the number of GMRES rounds that step consumed, as bars. The two residual curves coincide; the costs do not.}
    \label{fig:4-1-1}
\end{figure}

The solve itself uses \texttt{GVQSolver} with a subspace dimension of four, so that only $\log_{2}4=2$ qubits are used irrespective of the order of the system. The inner preconditioner is enabled at \texttt{prec\_lev}$=5$, and the backend is the exact state-vector simulator. The outer iteration is stopped when the largest power mismatch falls below $10^{-6}$ p.u.

To isolate the effect of the outer preconditioner, the same case is run a second time with that step reduced to rescaling and padding alone, that is, with \texttt{method='none'}. Both runs take exactly the same number of Newton steps, which is what makes the two comparable.

Figure~\ref{fig:4-1-1} shows both runs. The two residual curves lie on top of each other: each reaches the $10^{-6}$ p.u.\ tolerance at the sixth step, and at every step before that the two differ by less than the width of the line. What separates them is the right-hand axis. Over the six steps the unpreconditioned run consumes $2140$ GMRES rounds and \SI{22.7}{min} of time inside the solver calls; the run with the outer preconditioner consumes $7$ rounds and \SI{2.1}{s}, a factor of about $300$ in rounds and about $650$ in time.

The outer preconditioner therefore changes the price of the solve rather than its outcome: both runs walk the same trajectory, because a Newton step needs the direction of its correction far more than its agreement, and only the cost of each step separates them. This is a different role from that of the inner preconditioner of Section~\ref{sec:3-3}, which decided whether a solve worked at all; the two act at different levels and are not alternatives.

\begin{figure}[htb]
    \centering
    \includegraphics[width=\linewidth]{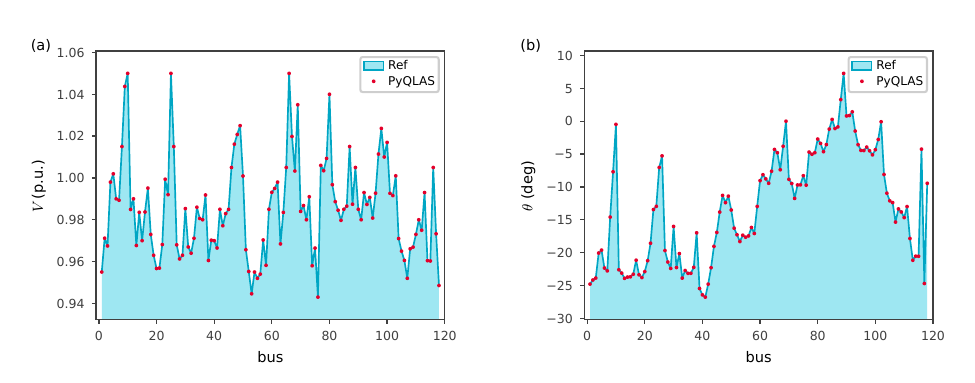}
    \caption{Solved voltage profile of the IEEE 118-bus system against the classical reference. (a) Voltage magnitude in per unit. (b) Voltage angle in degrees. The reference is drawn as a filled area and the solver's result as markers on its upper edge.}
    \label{fig:4-1-2}
\end{figure}

The solution itself is checked in Figure~\ref{fig:4-1-2}. Against a classical Newton--Raphson reference converged to $10^{-12}$, the largest deviation of the quantum--classical solve is $2.3\times10^{-8}$ p.u.\ in $V$ and $3.4\times10^{-6}$ degrees in $\theta$. These are the errors that the stopping rule of the outer iteration admits: at the last step the power mismatch is $1.6\times10^{-7}$ p.u., below the $10^{-6}$ p.u.\ at which the iteration stops, and the linear system solved at that step carries a relative residual of $3.4\times10^{-8}$. The quantum side therefore contributes no error beyond what the outer iteration itself allows. The profiles are those expected of this system: $V$ between $0.943$ and $1.050$ p.u., and $\theta$ spread over about $34$ degrees across the network.

The quantum resource consumed is small, and as in Section~\ref{sec:3-2} it is independent of the order of the system. Because the subspace dimension is four, the circuits use two qubits whether the block being solved is of order $181$ or $256$, and the whole solve consumed $2492$ circuits. The 181-order Jacobian never reaches the quantum device in any other form. The two times above are the elapsed time of the \texttt{GVQSolver} calls themselves---the same calls whose GMRES rounds are shown on the right-hand axis of Figure~\ref{fig:4-1-1}---and so cover both the machine time of Section~\ref{sec:3-5} and the library overhead around it. The classical work outside the calls---assembling each Jacobian, forming the mismatch, and applying the preconditioner---is additional: it is the price paid in classical time for the saving in quantum time, and at this size it is small, but it is not part of the \SI{2.1}{s}. Both runs were timed the same way, so the time factor of about $650$ compares like with like.

\section{Conclusion}\label{sec:5}

We have presented PyQLAS, a quantum--classical hybrid library for linear systems, and characterized it on synthetic systems and on one drawn from engineering practice. Its contribution is not another variational algorithm but a productized division of labour between the quantum device and the classical host. Three nested solvers---\texttt{VQSolver}, \texttt{IVQSolver}, and \texttt{GVQSolver}---share one variational core, so a user can move outward from a single solve to residual correction and then to a Krylov projection while the quantum subroutine keeps the same circuit width and the classical host absorbs the growth of the problem. Two independent levels of preconditioning, dimension alignment to a power of two, checkpointing and resumption, explicit device budgets, and a backend-independent interface complete the path from a matrix pair to a running hybrid solve. This combination distinguishes PyQLAS from single-layer VQLS prototypes and tutorials, from fault-tolerant linear-solver tools whose resource requirements are tied to long coherent evolution, and from application-specific products: it is a general library for the repeated linear solves that engineering workflows actually contain. On the IEEE 118-bus power flow, whose Newton Jacobians are of order $181$ and ill conditioned, the outermost solver reaches a maximum voltage-magnitude error of $2.3\times10^{-8}$ p.u.\ and a maximum angle error of $3.4\times10^{-6}$ degrees.

Three usage recommendations follow. First, prefer \texttt{GVQSolver} when quantum resources are the binding constraint: it confines the quantum work to a Krylov subspace of fixed dimension and inherits the robustness of the outer residual iteration, so the classical host can absorb problem growth. Second, enable the inner preconditioner whenever it is affordable: at $\kappa=1000$ it raises the success ratio of \texttt{IVQSolver} from $15\%$ to about $80\%$ and reduces the circuit count by roughly an order of magnitude; the outer preconditioner is an economy rather than a repair, because it changes the cost of the Newton path but not its direction, and its strength should be chosen with care since a coarse factorisation can leave the system worse conditioned than it began. Third, choose the backend and shot count to match the stage of the work: exact state-vector simulation for development, local sampling for noise-aware validation, and the cloud backend with modest shots for production, since raising shots does not reduce the number of circuits a solve consumes but does increase QPU time and cost. With these choices, the quantum part of an engineering computation can be small enough to be a design choice rather than a constraint, provided the classical side is allowed to do the work it is good at.

\section*{Acknowledgments}
This work was supported by the Anhui Provincial Science and Technology Innovation Tackling Program (Major Project) under Grant No.~202423s06050001, ``Research, Development and Application Demonstration of a Quantum Computing Cloud Platform Integrating Multi-System Quantum Computing Power with Supercomputing and Intelligent Computing Power''.

\printnomenclature

\NOMCLA{CNOT}{Controlled-NOT, a two-qubit quantum gate}
\NOMCLA{CPU}{Central processing unit}
\NOMCLA{CZ}{Controlled-Z, a two-qubit quantum gate}
\NOMCLA{GMRES}{Generalized minimal residual}
\NOMCLA{HEA}{Hardware-efficient ansatz}
\NOMCLA{LU}{Lower--upper, as in LU factorisation}
\NOMCLA{MVA}{Megavolt-ampere}
\NOMCLA{PQ}{Bus at which the injected active and reactive powers are specified}
\NOMCLA{p.u.}{Per unit}
\NOMCLA{PV}{Bus at which the injected active power and the voltage magnitude are specified}
\NOMCLA{QPU}{Quantum processing unit}

\NOMCLF{$A$}{coefficient matrix of the linear system}
\NOMCLF{$a$}{constant of the tridiagonal test family of Section~\ref{sec:3-3}}
\NOMCLF{$\B{b}$}{right-hand side of the linear system}
\NOMCLF{$c$}{constant of the tridiagonal test family of Section~\ref{sec:3-3}}
\NOMCLF{$\Delta\B{x}$}{correction applied to the state in one Newton step}
\NOMCLF{$\C{H}$}{a Hermitian operator, or the Hamiltonian it represents}
\NOMCLF{$I$}{identity operator}
\NOMCLF{$\B{J}$}{Jacobian matrix of the power-flow mismatch of Section~\ref{sec:4}}
\NOMCLF{$k$}{number of $R_Y$ layers in the ansatz family of Section~\ref{sec:3-4}}
\NOMCLF{$\kappa$}{condition number of a matrix}
\NOMCLF{$L$}{loss function of a solver, or its value}
\NOMCLF{$L_{\mathrm{mean}}$}{mean loss over a set of runs}
\NOMCLF{$\B{m}$}{vector of power mismatches in the Newton iteration of Section~\ref{sec:4}}
\NOMCLF{$M$}{dimension of the Krylov subspace, set by the \texttt{gmres\_dim} argument}
\NOMCLF{$N$}{dimension of the linear system}
\NOMCLF{$n$}{number of qubits, with $n=\log_{2}N$}
\NOMCLF{$N_c$}{number of quantum circuits consumed by a solve}
\NOMCLF{$N_\theta$}{number of parameters of an ansatz}
\NOMCLF{$P_s$}{success ratio over a set of runs}
\NOMCLF{$r$}{relative residual of a solution}
\NOMCLF{$R_\phi$}{parameterized phase gate, native to the processor of Section~\ref{sec:3-1}}
\NOMCLF{$R_Y$}{rotation about the $y$ axis of a qubit}
\NOMCLF{$T_1$}{energy relaxation time of a qubit}
\NOMCLF{$T_2$}{dephasing time of a qubit}
\NOMCLF{$\theta$}{parameters of the variational ansatz circuit (Figure~\ref{fig:2-1-1}), or phase angle at a bus (Section~\ref{sec:4})}
\NOMCLF{$V$}{voltage magnitude at a bus, of Section~\ref{sec:4}}
\NOMCLF{$\B{x}$}{solution vector of the linear system}

\bibliographystyle{elsarticle-num}
\bibliography{ref}
\end{document}